\documentclass[sigconf]{acmart}

\def\BibTeX{{\rm B\kern-.05em{\sc i\kern-.025em b}\kern-.08emT\kern-.1667em\lower.7ex\hbox{E}\kern-.125emX}}

\copyrightyear{2019} 
\acmYear{2019} 
\acmConference[CIKM '19]{The 28th ACM International Conference on Information and Knowledge Management}{November 3--7, 2019}{Beijing, China}
\acmBooktitle{The 28th ACM International Conference on Information and Knowledge Management (CIKM '19), November 3--7, 2019, Beijing, China}
\acmPrice{15.00}
\acmDOI{10.1145/3357384.3357906}
\acmISBN{978-1-4503-6976-3/19/11}

\usepackage{graphicx}
\usepackage{amsmath}
\usepackage{booktabs}
\usepackage{balance}
\usepackage{subcaption}
\usepackage{pifont}
\newcommand{\cmark}{\ding{51}}%
\newcommand{\xmark}{\ding{55}}%

\usepackage[ruled,linesnumbered]{algorithm2e}
\SetKwInOut{Input}{Input}\SetKwInOut{Output}{Output}

\usepackage{microtype}
\usepackage{booktabs,tabularx}

\begin{document}
\fancyhead{}
%
\title{Generating Persuasive Visual Storylines for Promotional Videos}

%
\author{Chang Liu}
\email{chang015@e.ntu.edu.sg}
\author{Yi Dong}
\email{ydong004@e.ntu.edu.sg}
\author{Han Yu}
\email{han.yu@ntu.edu.sg}
\author{Zhiqi Shen}
\email{zqshen@ntu.edu.sg}
\affiliation{%
  \institution{School of Computer Science and Engineering \\ Nanyang Technological University}
  \streetaddress{50 Nanyang Ave}
  \country{Singapore}
  \postcode{639798}
}
\author{Zhanning Gao}
\email{zhanning.gzn@alibaba-inc.com}
\author{Pan	Wang}
\email{dixian.wp@alibaba-inc.com}
\author{Changgong Zhang}
\email{changgong.zcg@alibaba-inc.com}
\author{Peiran Ren}
\email{renpeiran@gmail.com}
\author{Xuansong Xie}
\email{xingtong.xxs@taobao.com}
\affiliation{%
  \institution{Alibaba Group}
  \city{Hangzhou}
  \country{China}
}
\author{Lizhen Cui}
\email{clz@sdu.edu.cn}
\affiliation{%
  \institution{}
}
\affiliation{%
  \institution{School of Software \\ Joint SDU-NTU Centre for Artificial Intelligence Research (C-FAIR) \\ Shandong University}
  \city{Jinan}
  \country{China}
}
\author{Chunyan Miao}
\email{ascymiao@ntu.edu.sg}
\affiliation{%
  \institution{School of Computer Science and Engineering \\ Nanyang Technological University}
  \streetaddress{50 Nanyang Ave}
  \country{Singapore}
  \postcode{639798}
}

%
\renewcommand{\shortauthors}{Liu and Dong, et al.}

%
\begin{abstract}
        Video contents have become a critical tool for promoting products in E-commerce. However, the lack of automatic promotional video generation solutions makes large-scale video-based promotion campaigns infeasible. The first step of automatically producing promotional videos is to generate visual storylines, which is to select the building block footage and place them in an appropriate order. This task is related to the subjective viewing experience. It is hitherto performed by human experts and thus, hard to scale. 
        To address this problem, we propose WundtBackpack, an algorithmic approach to generate storylines based on available visual materials, which can be video clips or images. It consists of two main parts, 1) the Learnable Wundt Curve to evaluate the perceived persuasiveness based on the stimulus intensity of a sequence of visual materials, which only requires a small volume of data to train; and 2) a clustering-based backpacking algorithm to generate persuasive sequences of visual materials while considering video length constraints. In this way, the proposed approach provides a dynamic structure to empower artificial intelligence (AI) to organize video footage in order to construct a sequence of visual stimuli with persuasive power.
        Extensive real-world experiments show that our approach achieves close to 10\% higher perceived persuasiveness scores by human testers, and 12.5\% higher expected revenue compared to the best performing state-of-the-art approach.
\end{abstract}

%
%

%

%
\maketitle
    
    \section{Introduction}
    Over the years, E-commerce marketplaces, such as Alibaba and Amazon, are changing the way people shop and do business. Recently, there is an emerging trend in such platforms in which promotional videos increase the visibility and sales of products. However, it is still not very popular among the individual sellers or micro-sized enterprises because video production is usually a time-, skill- and cost-intensive process. This makes large-scale video-based product promotion campaigns on such platforms unfeasible.
    
    One way to address this issue is to create an intelligent system to automatically generate product promotional videos based on available product information (e.g. images, video clips, descriptions). Instead of focusing solely on the look and feel of the videos, we aim to generate videos that are persuasive. Persuasion is an attempt to influence a person's beliefs, attitudes, intentions, motivations, or behaviours, which have been studied in the fields of psychology \cite{okeefePersuasionTheoryResearch2016} and business \cite{armstrongPersuasiveAdvertisingEvidencebased2010}. In the context of the online marketplaces, persuasion aims to attract customers, make a good impression, and motivate purchases. \cite{armstrongPersuasiveAdvertisingEvidencebased2010}.
    
    In this paper, we focus on the first step of automatically generating promotional videos: given multiple visual materials (i.e. images or video clips) about a product, how to generate visual storylines with strong persuasiveness? This task involves selecting the visual materials and organizing them, while jointly considering three key aspects, Information, Attractiveness and Emotion. These considerations are very important for persuasive advertising \cite{armstrongPersuasiveAdvertisingEvidencebased2010}.

    Existing deep-learning-based video generation approaches require large amounts of training data. Since the perceived persuasiveness is highly subjective and may change with the category of products, such approaches must build very large video datasets and collect large numbers of feedbacks on each video, which is extremely time- and cost-intensive. 
    
    \begin{figure}
        \centering
        \includegraphics[width=1\columnwidth]{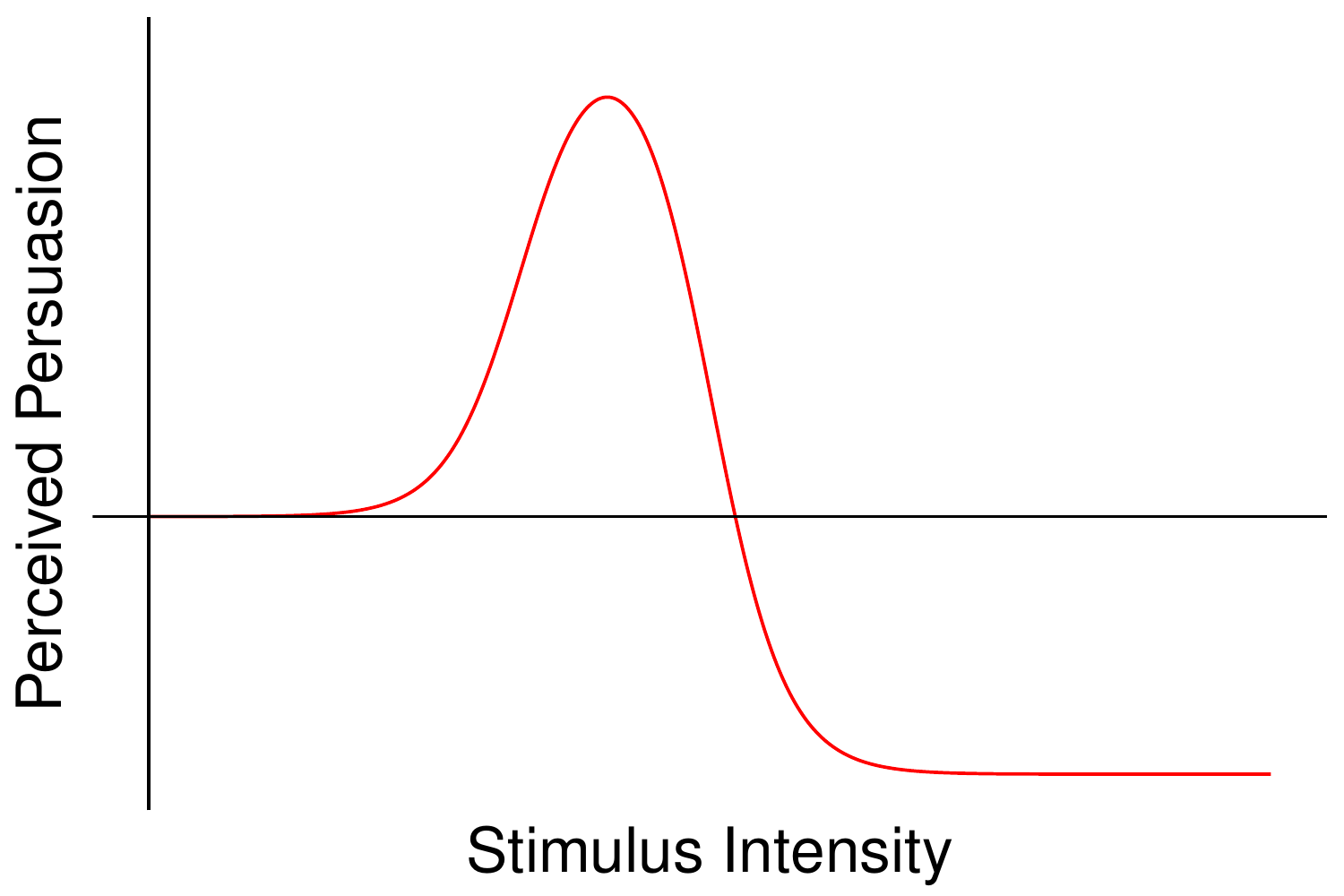}
        \caption{The Wundt Curve is an important curve in the filed of Psychology. It illustrates the relation between stimulus intensity and its effectiveness. We use the idea of the Wundt Curve to measure the perceived persuasion.}
        \label{fig:wundt}
    \end{figure}
    
\begin{figure*}
        \centering
        \includegraphics[width=0.8\linewidth]{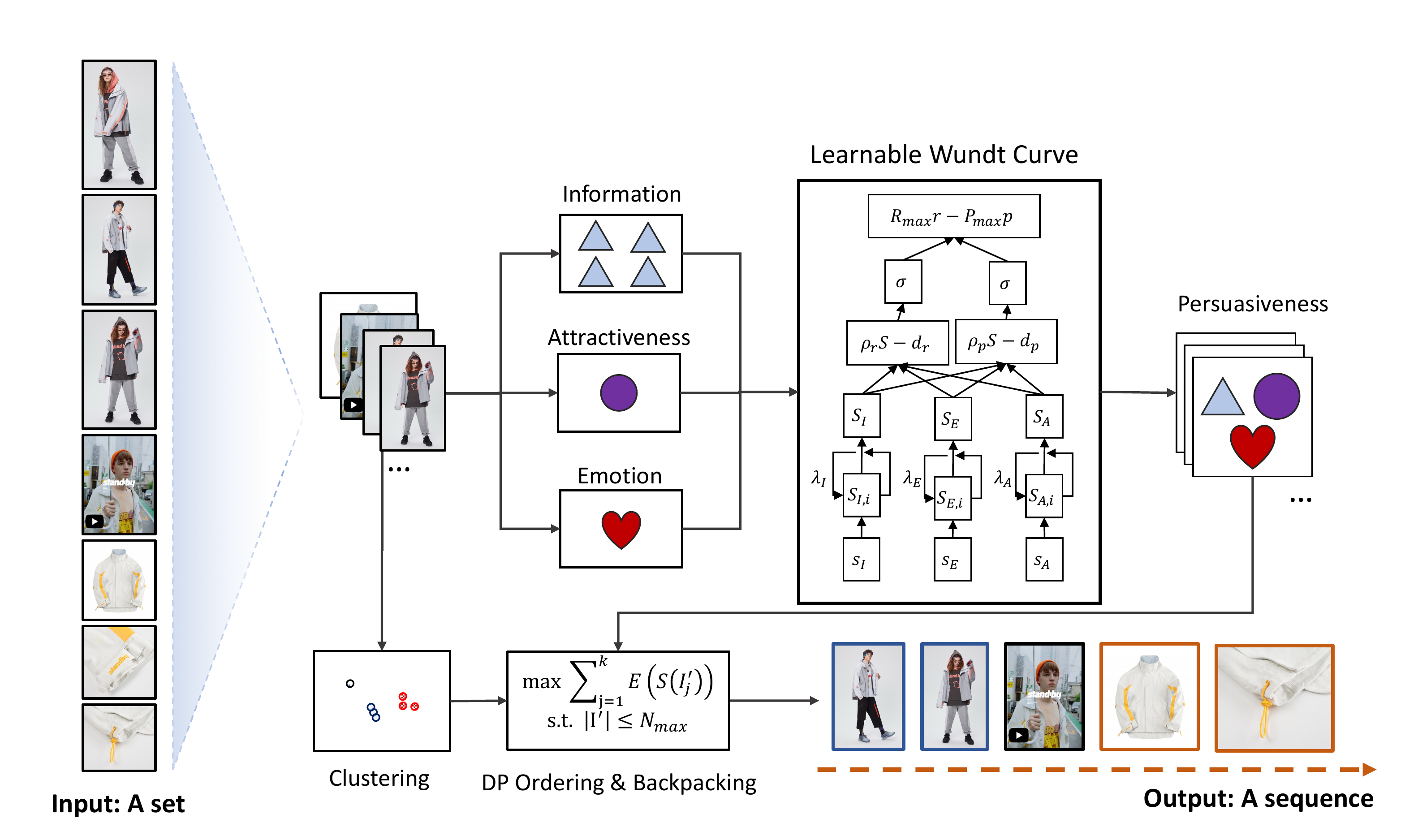}
        \caption{The proposed framework of WundtBackpack. The WundtBackpack takes a set of visual materials, which contains images or videos, as input, and outputs a visual storyline for persuasive promotional videos. It employs the Learnable Wundt Curve to predict the perceived persuasiveness based on the extracted features of information, attractiveness and emotion, and use clustering-based backpacking to do selection and sequencing.}
        \label{fig:arch}
    \end{figure*}

    In order to address this problem, we propose the WundtBackpack (WBP) algorithm, which is shown in Figure~\ref{fig:arch}. The first component is referred to as the Learnable Wundt Curve (LWC), which combines insights from Psychology with Machine Learning. The Wundt Curve proposed in \cite{berlyne1960} is a bell-shaped curve illustrating as the increase of stimulus intensity (shown in Figure~\ref{fig:wundt}), its effectiveness first increases to the highest then decrease rapidly. It has been widely used in the field of Psychology \cite{hirschman1980,cawthon2006,lepper2015,merrick2008}. LWC leverages machine learning approaches that quantify high-level features related to stimuli from a sequence of visual materials to learn the perceived persuasiveness. The combination of model-based approach with data-driven approach enables LWC to learn key Wundt Curve parameters with a small amount of data. The other component of WBP is Clustering-based Backpacking (CB) to compute the most persuasive visual storyline using dynamic programming.

    To better understand what videos can be persuasive, we build Persuasive Effectiveness Dataset, which consists of 20 promo videos of products from Taobao.com, with subjects' opinion towards informativeness, attractiveness, emotion and effects of persuasion of each video. To survey the persuasive effectiveness of videos, it is necessary to obtain enough feedback to reflect the general opinions. So each video has more than 30 pieces of feedback to ensure the averaged answer is unbiased. The detail of this dataset is given in Section~\ref{sec:dataset}.
    
    We evaluate WundtBackpack through real-world experiments to compare it against existing approaches. The results demonstrate that WundtBackpack achieves 9.9\% higher perceived persuasiveness scores by human testers, and 12.5\% higher expected revenue than the best state-of-the-art approach.
    
    To sum up, our contributions in this paper are as follows:
    \begin{itemize}
        \item We address the problem of automatically generating persuasive visual storyline for promotional videos. To our best knowledge, this is the first work on generating visual storylines that are persuasive. In this paper, we consider three persuasiveness principles: 1) Informativeness, 2) Attractiveness and 3) Emotion.
        \item We propose WundtBackpack algorithm to solve the problem of automatically generating persuasive visual storylines. It consists of the Learnable Wundt Curve technique, which is the component for evaluating the perceived persuasiveness considering stimulus intensity; and the Clustering-based Backpacking technique, which combines clustering and dynamic programming to identify the most persuasive sequence with regard to the score given by Learnable Wundt Curve, while ensuring the logic of the resulting visual storylines.
        \item We conduct extensive real-world experiments, which show that our approach achieves 9.9\% higher perceived persuasiveness scores according to human testers, and 12.5\% higher expected revenue compared to the best performing state-of-the-art approach.
    \end{itemize}
    
    The rest of this paper is organized as follows: Section 2 reviews the previous work on automatic video generation as well as visual material sequencing. Section 3 introduces how a promotional video, especially its visual storyline, shows persuasiveness; and explains the features adopted by our system. In Section 4, our proposed algorithm, the WundtBackpack (WBP) is explained in detail. Section 5 introduces our real-world experiments and provides analysis of the experimental results. Finally, in Section 6, we conclude our paper and discuss possible future research directions.

    \section{Related Work}
    Promotional video generation is a complex task which is composed of content analysis, storyline generation, visual effects generation \cite{hua2006photo2video}. Several attempts have been made on these tasks. 
    \citeauthor{hua2006photo2video} proposed the ``Photo2Video'' system \cite{hua2006photo2video} to generate a video story. It takes the collection of images and a piece of music as input. The system identifies the key actor from the images and uses it to generate the sequence of images. The visual effects are created by framing scheme with consideration of music rhythm jointly. However, the storyline is defined by humans.
    
    \citeauthor{wu2016monet} presented an improved system of ``Photo2Video'' called ``Monet'' \cite{wu2016monet} which considers event segmentation and photo selection. Both of them \cite{hua2006photo2video,wu2016monet} assumes the existence of the time and location (e.g., GPS information) that photo is taken, and use them to do photo selection, grouping and sequencing. However in the real-world E-commerce platform, such assumptions are less likely to be true since the sellers always edit, crop and concatenate photos before put it online, also the E-commerce platform will delete these meta-data to minimize the size of the file. All of them made the date and location of images very hard to retrieve. In addition, they can not accept video clips as input.
    
    To solve the problem of visual material selection and ordering, we then investigate the visual material sequencing methods, who focus on analyzing contents instead of the meta-data of visual materials.
    
    \subsection{Visual Material Sequencing}
    Visual material sequencing aims to generate an order of images or videos in order to achieve some design objectives. Using machine learning to perform visual material sequencing has been studied in several recent research. Approaches include sorting image-caption pairs \cite{agrawal2016,wang2012generating}, composing video-stories \cite{choi2016,zhong2018}, and optimizing image ordering \cite{sigurdsson2016}. A comparison between our proposed approach and existing work can be found in Table~\ref{tab:pre_comp}.
    
        
    \begin{table}[ht]
        \centering
        \setlength\tabcolsep{2.7pt}

            \centering
            \begin{tabular}{lcccc}
                \toprule
                & SRNN & PA & 2RNNS  & WBP \\    
                \midrule
                Able to take images as input            & \cmark            & $\circ$                 & $\circ$                                        & \cmark                 \\
                Able to input video clips       & \xmark            & \cmark                 & \cmark                                        & \cmark                 \\ 
                Able to select visual materials            & \cmark            & \xmark                 & \xmark                                        & \cmark                 \\ 
                Accounts for attractiveness      & \xmark            & \cmark                 & \cmark                                        & \cmark                 \\ 
                Accounts for logic            & \xmark            & \xmark                 & \xmark                                        & \cmark                 \\ 
                Accounts for informativeness            & \cmark            & \cmark                 & \cmark                                        & \cmark                 \\
                Accounts for emotion                    & \xmark            & \xmark                 & \xmark                                        & \cmark                 \\ 
                Performs well with few data & \xmark            & N/A             & \xmark                                        & \cmark       \\
                \bottomrule         
            \end{tabular}
            \captionof{table}{Comparison between the proposed approach and existing solutions, Skip-RNN (SRNN) \cite{agrawal2016}, Plot Analysis (PA) \cite{choi2016} and Two Stream RNN with Submodular Ranking (2RNNS) \cite{zhong2018}. $\circ$ means the function is originally not supported, but could be supported with minor adjustments.}\label{tab:pre_comp}
        \hfill
        \end{table}
    
    \citeauthor{agrawal2016} addressed the problem of sorting jumbled image-caption pairs into stories \cite{agrawal2016}. The idea is to find the temporal order of events by analyzing both images and captions. They proposed several approaches to solve this problem, and the best algorithm has been found to be the ensemble method. It uses LSTM \cite{hochreiter1997} to understand captions and employ GRU \cite{cho2014} with deep features of images to give another order with regard to image-caption pairing. Then, the optimal permutation is computed using a voting scheme.
    
    \citeauthor{sigurdsson2016} uses Skip RNN to skip through certain images via a sampling approach, in order to sequence a set of photos \cite{sigurdsson2016}. The skipping is performed by introducing latent variables into RNN models. They collect Flicker albums dataset to train and evaluate algorithms.
    
    The above-mentioned works \cite{sigurdsson2016,agrawal2016} take temporal order as the key sorting criterion. However, considering temporal order alone may not optimize the viewing experience. \citeauthor{choi2016} proposed the problem of video-story composition, which is to sequence a set of video clips considering semantics, motions and dynamics \cite{choi2016}. They calculate dissimilarity and the plot penalty, which encourages the sequence of video clips to follow the proposed general plot of stories.
    
    \citeauthor{zhong2018} presented an improved solution for video-story composition \cite{zhong2018}. They used two-stream RNN, which takes C3D \cite{tran2015} and SPP-HOOF (Histogram of dense optical flow with spatial pyramid pooling) feature as input, and learns to pick the next clip based on previously selected visual materials. They also proposed a sub-modular ranking method to refine the results produced by RNN.
    
    From the above review, we can observe that none of the related work is able to sequence video clips and images at the same time. In addition, the criteria (or features) used for sequencing mainly involve informativeness and attractiveness. The effect of emotion is rarely considered. This makes them hard to apply in marketing. Also, data-driven approaches \cite{agrawal2016,sigurdsson2016,zhong2018,yi19personalized} require large amounts of data. The proposed WundtBackpack approach addresses these limitations simultaneously.
    
    \section{What Makes Videos Persuasive?}\label{sec:feature}
    \citeauthor{armstrongPersuasiveAdvertisingEvidencebased2010} introduced a few successful principles and tactics in persuasive advertising \cite{armstrongPersuasiveAdvertisingEvidencebased2010}. In the context of persuasive videos, three key aspects should be considered:
    \begin{itemize}
        \item \textbf{Informativeness} measures how much useful information in videos relates to products, which includes performance, features, prices, etc. In addition, the logic of presentation regarding the information is important for successful persuasion.
        \item \textbf{Attractiveness} motivates the customers to take action. For example, using fashion models to enhance the look and feel of clothes, thus attract the viewers.
        \item \textbf{Emotion} can affect consumer expectations, which in turn can affect satisfaction derived from purchasing. Some emotional tactics used in the field of advertising include trust, self-expression and fear.
    \end{itemize}
    To quantify the above-mentioned key aspects of persuasion, we propose three features to represent visual materials.
    
    \textbf{Dissimilarity}. Generally, most of the product introduction pages in E-commerce platforms contain sufficient information for making purchasing decisions \cite{laudon2017commerce}. Therefore, we do not need to consider what information does the visual materials must cover and how it relates to the product. We only need to keep the information which is non-redundant. In this way, we propose to use dissimilarity of two adjacent visual materials to quantify information. In this paper, Structural Dissimilarity \cite{wang2004} is employed.

    \textbf{Aesthetics}. To compute a score on attractiveness, we choose to employ the aesthetics score, which measures how beautiful a visual material looks to humans. Here, we leverage previous research on designing deep learning models to predict the aesthetics scores of images \cite{deng2017,schwarz2018,talebi2018,wang2018}.
    
    \textbf{Arousal}. The induced emotion when users view visual materials can be characterized by a 2-dimensional model of valence and arousal \cite{russell1980}. Arousal measures the level of emotional stimuli, and valence measures whether the emotion is positive or negative. Here we only employ arousal to give scores on the emotion because studies show a high correlation of media memorability and arousal \cite{cahill1995novel} while it is hard to determine whether it is more persuasive to use positive or negative emotion in different categories of product.

    Note that the ``logic'' in the information aspect is not represented as a feature. Later in section~\ref{sec:cb} we will show how it is represented in our algorithm.
    
    \section{The WundtBackpack Approach}
    The proposed WundtBackpack algorithm is composed of two parts, 1) Learnable Wundt Curve (LWC) to evaluate the perceived persuasiveness of the sequence of visual materials, and 2) Clustering-based Backpacking (CB) to find the optimal sequence with regard to the persuasiveness score. Here, we want to maximize the perceived persuasiveness:
    \begin{align}\label{equ:obj}
    \begin{split}
    \text{Max: } &\sum_{j=1}^{k} E(S(I'_j)), \\
    \text{s.t. } &|I'|<=N_{\max}.
    \end{split}
    \end{align}
    The decision variable is $I'=[I'_1,...,I'_k]$. Each element $I'_j$ refers to the sequence of the selected visual materials in cluster $V_j$. There are overall $k$ clusters of visual materials. $N_{\max}$ is the maximum number of visual materials. $E(x)$ represents the perceived persuasiveness of stimulus intensity $x$, and $S(I')$ refers to stimulus intensity of sequence $I'$, which will be introduced in more details later.
    
    The reason for clustering is that 1) Permutation algorithm has very high time complexity. Doing permutations on the whole set $V$ is time-intensive. However, by carefully selecting $K$ value, the time for permutation on each cluster is acceptable. The method of choosing $k$ value will be introduced in Section~\ref{sec:cb}. 2) The similar visual material usually represents a similar topic. Therefore by arranging them to be together will incorporate the logic into the visual-storylines. Here, ``logic'' refers to introducing one part after another, e.g., providing an overview of the product before zooming into an interesting part. It is crucial for making the structure of videos meaningful. 
    
    \subsection{Learnable Wundt Curve (LWC)}
    LWC evaluates the persuasiveness of given stimulus intensity.
    
    In the field of computer science, research in curiosity agents \cite{wu2013} employed the Wundt Curve to estimate the effectiveness of interestingness \cite{saunders2002} and novelty \cite{merrick2008}. In \cite{saunders2002}, the Wundt Curve is formulated as the difference of two sigmoid functions, reward and punishment:
    \begin{align}\label{equ:wundt}
    \begin{split}
    R(x)=\frac{R_{\max}}{1+e^{-\rho_R(x-d_R)}}, \\
    P(x)=\frac{P_{\max}}{1+e^{-\rho_P(x-d_P)}}, \\
    E(x)=R(x)-P(x),
    \end{split}
    \end{align}
    where $R_{max}$, $\rho_R$ and $d_R$ are the maximum value of the reward, the slope of the reward function, and the minimum stimulus to be rewarded, respectively. The parameters of punishment $P_{\max}$, $\rho_P$ and $d_P$ are similarly defined.
    
    The challenge of Wundt curve formulation is that there are 6 parameters to be quantified. One can observe the function value and tune the parameters by hand. However, it requires a lot of time and domain knowledge. Once the context is changed (e.g. different categories of products), tuning needs to be performed again. In order to address the manual tuning limitation, we propose LWC, which requires a small amount of data in order to produce an accurate Wundt Curve.
    
    In LWC, we employ the same formulation as in \cite{saunders2002}, as shown in Equation~(\ref{equ:wundt}). Training is performed by gradually moving the parameters towards the minimum of the cost function. For example, to optimize the slope of the punishment function $\rho_P$, it can be trained as
    \begin{align}
    \rho_P = \rho_{P0}-\alpha\triangle\rho_P
    \end{align}
    where $\alpha$ is a fixed learning rate.
    
    To calculate the gradients, we employ the chain rule,
    \begin{align}
    \begin{split}\nonumber
    \triangle\rho_P=\frac{\partial L}{\partial \rho_P}&=\frac{\partial L}{\partial E}\frac{\partial E}{\partial Z}\frac{\partial Z}{\partial \rho_P}
    =\frac{(x-d_P)e^Z}{(e^Z+1)^2}l'(x) \\
    &= E(x)(1-E(x))(x-d_P)l'(x),
    \end{split}
    \end{align}
    \begin{equation}\nonumber
    \triangle d_P=\frac{\partial L}{\partial d_P}= l'(x)E(x)(1-E(x))\rho_P
    \end{equation}
    \begin{align}
    \begin{split}\nonumber
    \triangle P_{max}=\frac{\partial L}{\partial P_{max}}&=\frac{l'(x)}{1+e^{-Z}}
    \end{split}
    \end{align}
    where $Z=\rho_P(x-d_P)$ and $L'(x)$ is the first derivative of the loss function. Similarly we define the gradients of other parameters $\rho_R, d_R, d_R, R_{\max}$.
    
    In this way, the Wundt curve can approach to the position where the loss function is minimized with regard to the given data. However, since the Wundt curve measures the relation between stimulus intensity and perceived persuasiveness, it is still unclear what the stimulus intensity is composed of. In other words, there are gaps between features of visual material and stimulus intensity of a given visual material sequence. Here, we use a structure similar to RNN to bridge the gap.
    
    As for dissimilarity $s_{d_i}$, aesthetics $s_{a_i}$ and emotional arousal $s_{e_i}$ of the $i$-th visual material in $I'$, the stimulus intensity $S$ of the visual material sequence $I'$ is expressed as
    \begin{align}
    x=S=w_d*S_d+w_a*S_a+w_e*S_e
    \end{align}
    where
    \begin{align}
    \begin{split}
    S_d&=S_{d,n}=W_d \cdot a_n+\lambda_d S_{d,n-1}.
    \end{split}
    \end{align}
    
    Since we aim to design LWC to use very little data to learn the Wundt Curve paramters, the stimulus intensity and features of images we employed are scalars. Thus we can remove the parameter $W$ (i.e. by setting $W=1$):
    \begin{equation}
    w_d*S_d=w_d (W_d a_n+\lambda S_{d,n-1})=w'a_n+\lambda' S_{d,n-1}.
    \end{equation}
    Similar with the method for calculating $\triangle \rho_P$, the gradient of $\lambda_d$ is,
    \begin{align}\nonumber
    \begin{split}
    \triangle\lambda_d=l'(x) (\rho_p+\rho_r) w_d (a_{n-1} +2\lambda a_{n-2} + \\
    ...+ (n-1)\lambda^{n-2} a_1).
    \end{split}
    \end{align}
    
    In this way, the function $S()$ can be trained together with $E()$, thus, in the rest of this paper, we will use $E(I')$ to represent the learned function mapping features of visual materials to perceived persuasiveness.
    
    \subsection{Clustering-based Backpacking (CB)}\label{sec:cb}
    In this subsection, we will introduce Clustering-based Backpacking, which utilizes the model produced by LWC, to find the sequence $I'$  that optimizes the objective function in Equation (\ref{equ:obj}). The pseudo code of CB is given in Algorithm~\ref{algo}.
    
    \begin{algorithm}
        \SetAlgoLined
        \Input{$V$: the set of visual materials\\$k$: the number of clusters\\$N_{\max}$: maximum number of visual materials\\ $E()$}
        \Output{$I'$: a sequence of the visual materials in $V$ with the maximum perceived persuasiveness}
        \BlankLine
        $[V_1,V_2,...,V_k]$= K-means($V$,$k$) \\
        $BS=$ empty 2-d array \\
        \ForEach{$j \in [1,k]$}{
            $BS[j][i]$= best sequence with every possible length $i$ w.r.t. $E()$ ($i \in [1,|V_j|]$)
        }
        $dp$ = results of dynamic programming produced by Equation (\ref{equ:dp})\\
        $score$ = the best score in $dp[k]$ considering the objective function in Equation (\ref{equ:obj}) \\
        $I'$ = the sequence of visual materials achieving the $score$
        \BlankLine
        \caption{Clustering-based Backpacking}\label{algo}
    \end{algorithm}
    CB first uses K-means \cite{macqueen1967} to cluster the set of visual materials. Then, for each possible length of visual materials in each cluster, it finds the best sequence with regard to $E()$. For the K-means clustering, we use Euclidean distances on the deep feature extracted from the fc7 layer of AlexNet \cite{krizhevsky2014}. For videos, the distance is calculated by the minimum distance of the image and each frame. The number of clusters ($K$) is defined explicitly. For different categories of products, $K$ can be different to achieve the best viewing experience. It should be defined with regard to the general structure of introduction for a certain category of products, i.e., $K$ = generally how many sections are in the introduction for this category. For example, in the context of selling clothes, $K=3$ is recommended since most of the clothes introductions are composed of three parts, model trying-on, detail showing and other information.
    
    In this way CB computes a set of ``grouped items'', each item (sequence $I'$) is assigned a weight (length), belongs to a group (cluster), and has a value $E(I')$. This forms the next step of the algorithm as the following question: given a set of ``grouped items'', with maximum weight $N_{\max}$, and each group has at most 1 item to be selected, how to find a combination of items to maximize the objective function in Equation (\ref{equ:obj}).
    
    This problem can be solved by dynamic programming. Here we set the state of the dynamic programming model as $dp(j,i)$, which means that the maximum value of the objective function (\ref{equ:obj}) considering that the first $j$ groups have exactly $i$ weights. For one state $dp(j,i)$, it can be transferred from either $dp(j-1,i)$, which means for the $j$-th group no item is selected, or $dp(j-1,i-l)$, which means the item weighted $l$ in $j$-th group is selected. It can be observed that such design of states has an optimal substructure. Mathematically, the state transition equation is given as follows,
    \begin{align}\label{equ:dp}
    \begin{split}
    dp(j,i) &=\max  \{ dp(j-1,i),  \\
    &dp(j-1, i-l) + E(BS[j][l]) \text{ }| l \in [1,|V_j|]\}.
    \end{split}
    \end{align}
    Note that, in Equation (\ref{equ:dp}), only $dp(j-1,)$ is used for computing $dp(j,)$. Therefore, in the implementation, the state can only be stored in a 1-dimensional array $dp(i)$. The final decision state is the state who has maximum value. In addition, The corresponding sequence of $dp(i)$ shall be stored in order to avoid backtracking to improve search efficiency.
    
    \section{Experimental Evaluation}
    We conduct real-world experiments to compare the performance of our approach against existing solutions.
    \subsection{Experiment Settings}
    The 6 comparison approaches are:
    \begin{itemize}
        \item RNN: which takes dissimilarity, aesthetics and arousal of each visual material in sequence as input, and predicts the perceived persuasiveness of the sequence. The baseline model is composed of a 1-layer RNN followed by a fully connected layer. 
        \item LSTM: in which forget gate, input gate, and output gate are added to the RNN model to form an LSTM \cite{hochreiter1997}. The settings are the same as RNN model.
        \item GRU: which combines the forget and input gates into a single ``update gate'' \cite{cho2014}. The input, output and model structures are the same as RNN.
        \item Plot Analysis (PA) \cite{choi2016}: which calculates the dynamics and dissimilarity to search for the best order of all visual materials. In the experiment, we only keep the first $N_{\max}$ visual materials.
        \item Two Stream RNN with Sub-modular Ranking (2RNNS) \cite{zhong2018}: which uses 2 RNNs and a sub-modular ranking considering C3D and SPP-HOOF features. We obtain the C3D feature of an image by duplicating the image 16 times and sending them to the C3D model. Same as PA, only the first $N_{\max}$ visual materials are preserved.
        \item Skip-RNN (SRNN) \cite{sigurdsson2016}: which skips through certain images via latent variables. It is not able to select videos and images at the same time. Thus, during experiments, videos are not used by SRNN.
    \end{itemize}
    We use Pytorch \cite{paszke2017automatic} to implement LWC. The aesthetics score of visual materials is extracted by the NIMA pretrained model \cite{talebi2018}. As for arousal, we adopt MobileNet \cite{howard2017} and train the network on the Image Emotion Dataset \cite{kim2018} to extract the information. We extract each frame from a video and compute the maximum scores for dissimilarity, aesthetic, and arousal. In addition, a dynamic incentive \cite{choi2016} is added to the attractiveness of the videos.
    
    Since all the above-mentioned methods do not consider the duration of presenting each image, we set it to be the same for all images (1.5 seconds in our experiments), and the video clips are played at a normal speed. $N_{\max}$ is set as 8.
    \subsubsection{The Persuasion Effectiveness Dataset}\label{sec:dataset}
    We collected a dataset called the Persuasion Effectiveness Dataset (PED), which is used to train our model. We obtained 20 product introduction videos from Taobao and Tmall. The categories of products include pants, upper body clothing, accessories, shoes and skirts. These videos are no longer than 60 seconds. They are carefully selected by domain experts to cover frequently used video techniques (e.g. multiple footages, single footage, slides). Domain experts assessed that half of the videos have good persuasiveness while others do not. The principles of persuasiveness are given in Section~\ref{sec:feature} and \cite{armstrongPersuasiveAdvertisingEvidencebased2010}. Then, we build a website to collect user opinions about the perceived persuasiveness of the videos. We employ a 5-point Likert scale. We asked about the information, attractiveness, emotion and perceived persuasiveness of sequence. Each feedback (rated score) for a question is mapped to an integer value in the range of$[0,4]$. After collecting enough data on a video, the data are averaged to reflect the general opinion of all subjects.
    
    \begin{figure*}[h]
        \centering
        \includegraphics[width=1\linewidth]{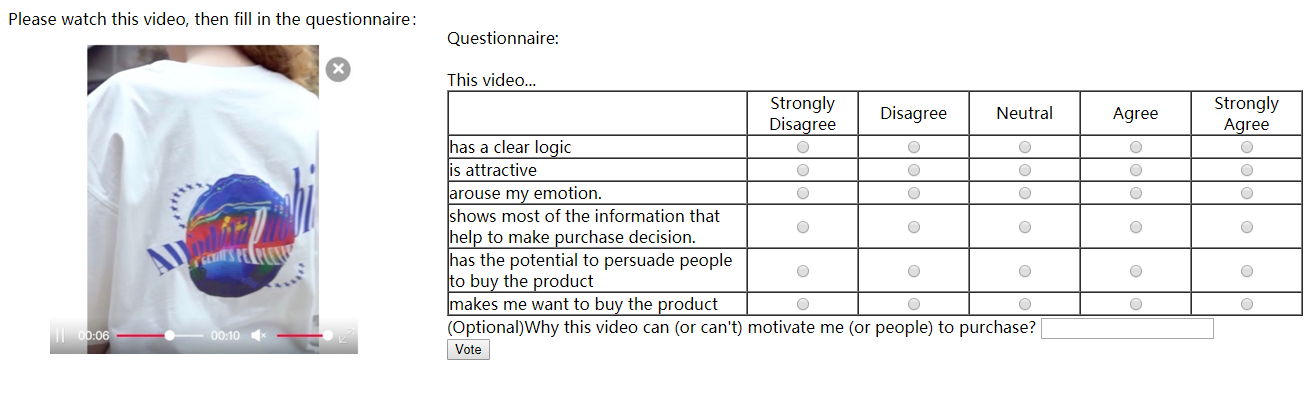}
        \caption{The screenshot of the user opinions collection website. Subjects are asked to fill up a questionnaire about the video they just watched. They may skip parts (or all) of the questions if they want.}
        \label{img:website}
    \end{figure*}
    In Persuasive Effectiveness Dataset we design two similar but different questions, ``The video has the potential to persuade people'' and ``The video makes me want to buy this product''. We want to see whether the viewers will hold different opinions towards their own perceived persuasiveness and the persuasive effectiveness of the general public. In our dataset, we find that the average respond of ``persuade people'' on all the videos is 2.093, while for  ``persuade me'' is 1.791. The difference is not large. However, this shows that when considering the persuasiveness effectiveness of the general public, the viewer tends to think in a more optimistic way. In addition, the average responds prove the assertion which half of the videos are good.
    
    During opinion collection, the order of presenting the videos to a viewer may affect the results. For example, after seeing a very persuasive video, the next video, which has middle-level persuasiveness, may receive a very low score due to the anchoring effect. To address this problem, the video display order is randomized for each viewer, and there are 33 participants involved overall.
    
    We performed a regression testing between the information, attractiveness, emotion and the perceived persuasiveness. A linear SVM for regression \cite{chang2011} with $C=1.0, \epsilon=0.1$ is employed. The results are shown in Table~\ref{tab:results_principle}. The three proposed aspects and the perceived persuasiveness have a very strong positive correlation ($R=0.8902$). From our collected data information is the most important aspect for perceived persuasiveness, which accounts for an almost 45\% weights followed by attractiveness (about 42\% weight), and the emotion (about 13\% weight). We think the reason for the low weightage of emotion is that it is not an important factor affecting purchasing decisions for clothing. Note that these results are only valid in the context of the E-commerce platform with the selected categories of products. Once the context is changed, the weights may also be different.
    \begin{table}[]
        \centering
        \begin{tabular}{lr}
            \toprule
            Attributes                          & Values             \\ 
            \midrule
            Correlation coefficient $R$                                 & 0.8902 \\ 
            Weights assigned to Logic (Information)           & 0.3243         \\ 
            Weights assigned to Informativeness & 0.1265         \\ 
            Weights assigned to Emotion         & 0.1260         \\ 
            Weights assigned to Attractiveness  & 0.4241        \\ 
            \bottomrule
        \end{tabular}
        \caption{The regression results on the PED}
        \label{tab:results_principle}
    \end{table}

    The dataset used to evaluate the generated visual story-line is composed of 20 sets of visual materials. Thirteen of them contain videos. The average number of visual materials is 14. They are obtained from Taobao and Tmall, with the same product categories as those in the PED. Note that there is no overlap between the evaluation data and the training data, PED.

    \subsection{Evaluation Metrics}
    Firstly, we train our models on PED. Our proposed LWC model is compared against baseline models RNN, LSTM and GRU. The Mean Squared Error (MSE) is selected as the evaluation metric. Note that we do not split data into the training and testing sets. This is because, 1) the dataset is small, and splitting it will make the training (or testing) set not covering all the types of videos; 2) since we only have 12 parameters, and the goal of training is only to learn the key parameters for the Wundt Curve. The shape of LWC is fixed. Therefore, there is a low possibility of overfitting. While for the traditional NN model, such a problem may exist.
    
    Secondly, we evaluate the WundtBackpack approach and state-of-the-art visual material sequencing algorithms by pairwise comparison. The pairwise comparison is performed as follows: in every round, two videos generated from the same inputs but different algorithms will be presented to a human subject at the same time. The user interface is similar to Figure~\ref{img:website}, but presents two videos and only one question in the questionnaire. We do not tell the subjects which algorithms the videos are generated with. The subject makes a choice on which video has a better perceived persuasiveness after watching the two videos, and they are allowed to replay the videos. A video rated as  ``significantly better'' by a subject receives a score of 2, while a video rated as``better'' will receive a score of 1. The scores are then averaged over the number of times the video has been displayed to subjects. Subjects may skip a question if they believe the two videos are equally persuasive.
    
    Four algorithms, PA, 2RNNS, SRNN and WBP are compared in the pairwise comparison. The choice of algorithms and the sequence of the videos are randomized. The average score for each video represents the final perceived persuasiveness score. In this way, the bias is minimized. Overall we engaged 67 subjects, and they performed 1,103 pairwise comparisons.

    \subsection{Results and Analysis}
        \begin{figure}
        \centering
        \begin{minipage}[h]{0.9\columnwidth}
            \centering
            \includegraphics[width=\linewidth]{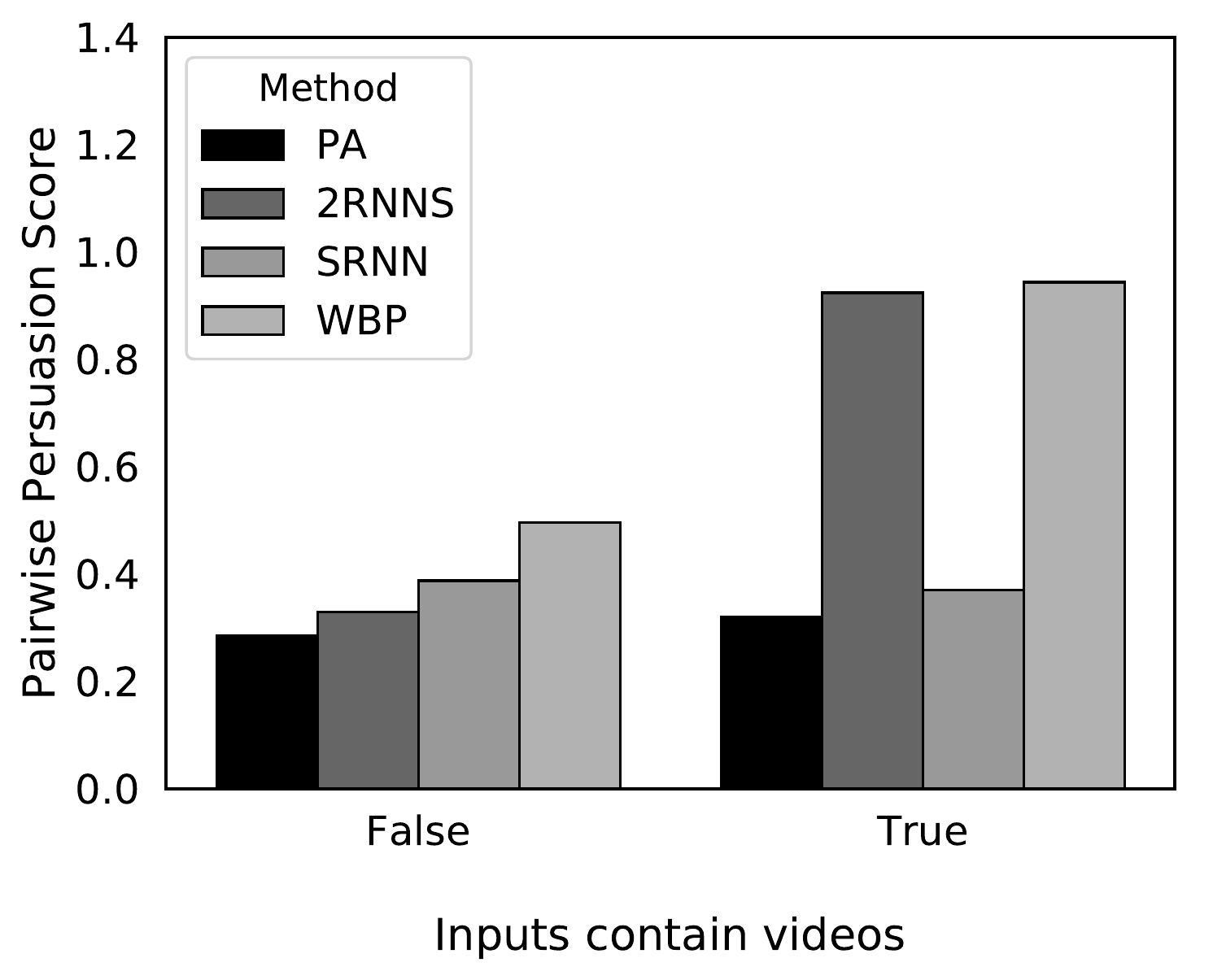}
            
            \caption{The pairwise persuasion scores of inputs containing video and not containing}
            \label{img:wo}
        \end{minipage}\hfill
        \begin{minipage}[h]{0.9\columnwidth}
            \centering
            \includegraphics[width=\linewidth]{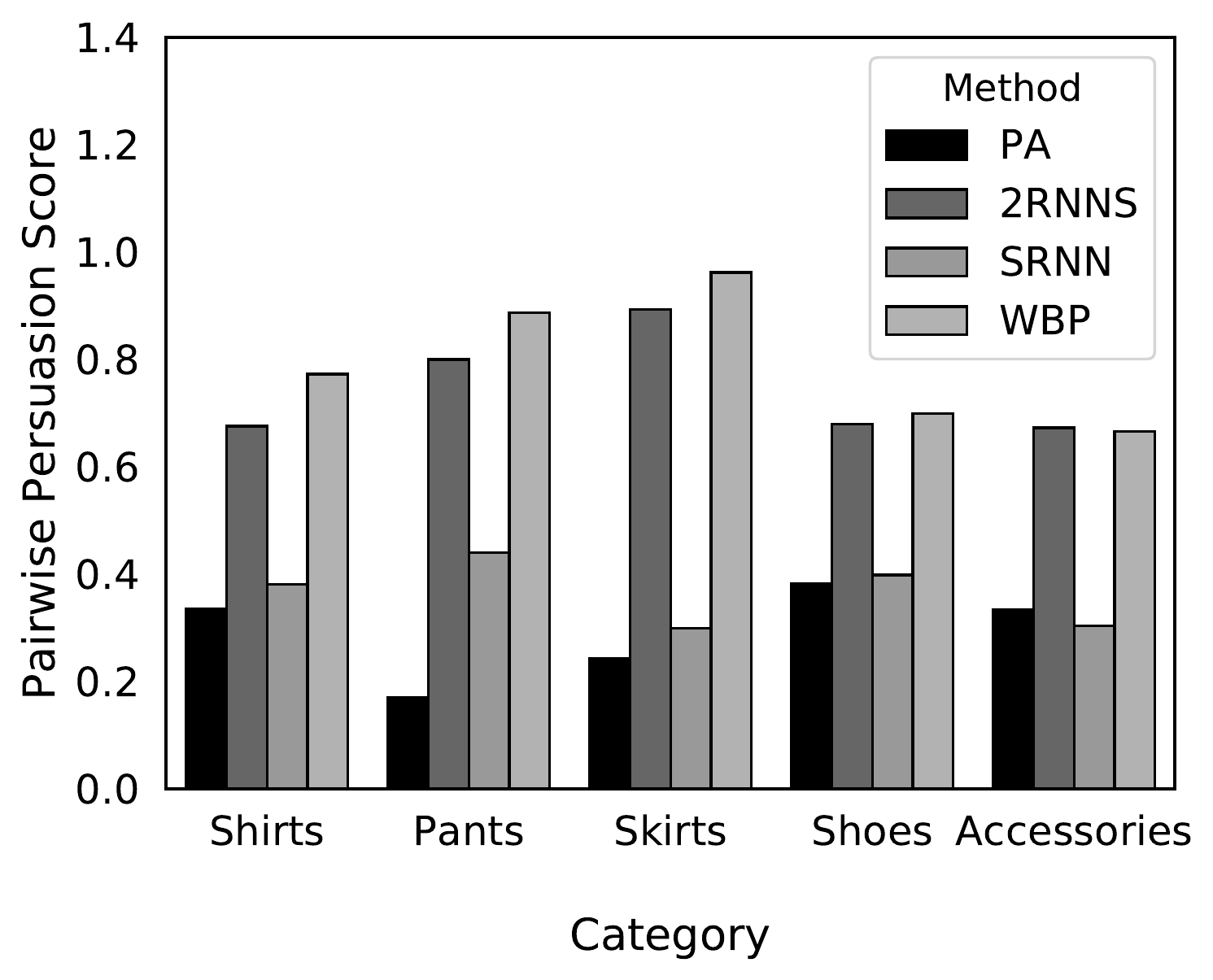}
            \caption{The pairwise persuasion scores among different categories of products}
            \label{img:cat}
        \end{minipage}\hfill
        \begin{minipage}[h]{0.9\columnwidth}
            \centering
            \includegraphics[width=\linewidth]{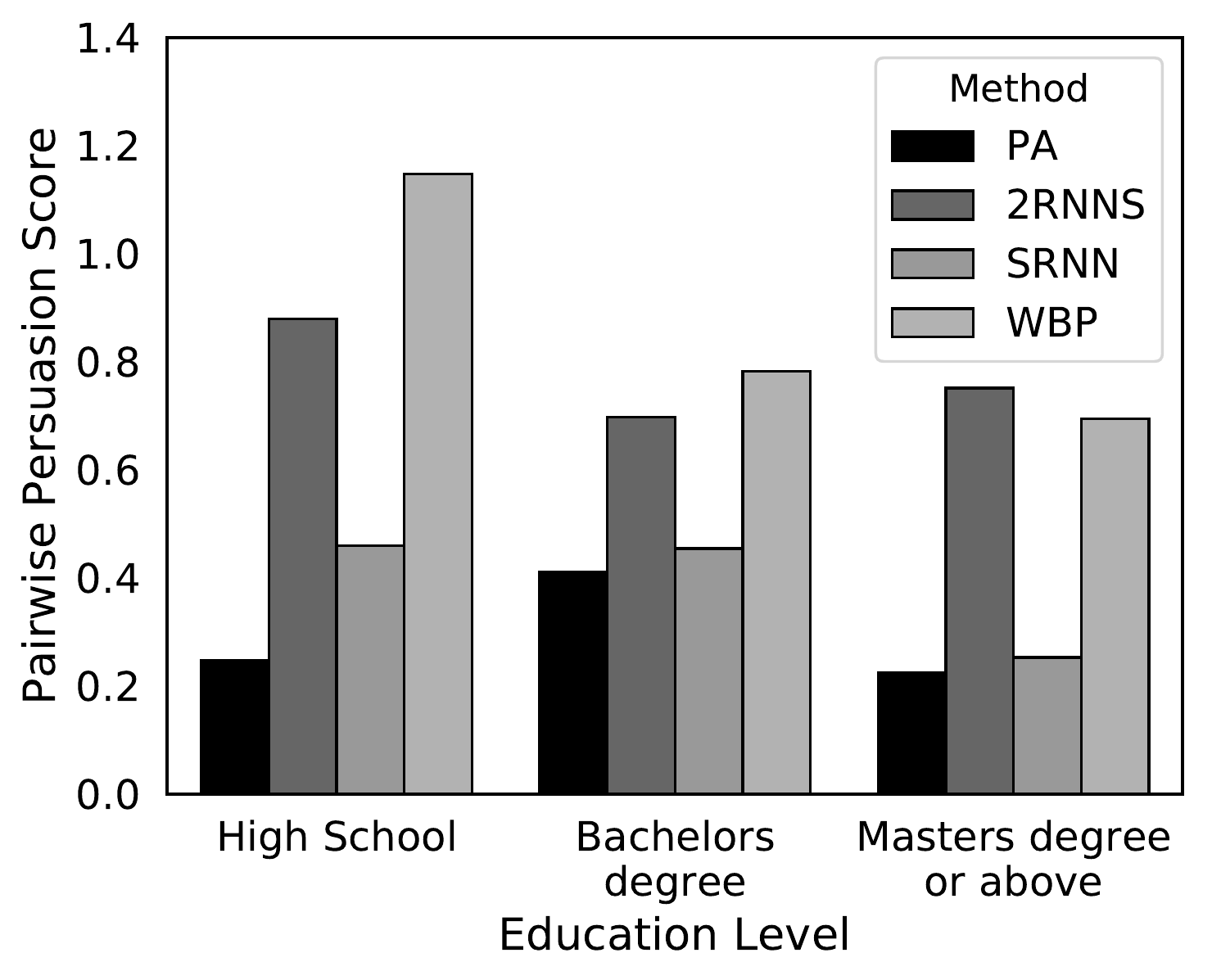}
            \caption{The pairwise persuasion scores for users of different educational levels}
            \label{img:edu_all}
        \end{minipage}\hfill
    \end{figure}

    \begin{table}[]
        \centering
        \begin{tabular}{lrr}
            \toprule
            Model     & \# Params. & MSE Loss \\
            \midrule
            RNN       & 28           & 0.15573  \\
            LSTM      & 100          & 0.15524 \\
            GRU       & 76          & 0.14656 \\
            LWC, Ours & 13           & \textbf{0.11657} \\
            \bottomrule
        \end{tabular}
        \caption{The training results of proposed LWC and its baselines}
        \label{tab:training_result}
    \end{table}

\begin{table}[]
\begin{tabular}{@{}ll@{}}
\toprule
Features & MSE Loss         \\ \midrule
Arousal       & 0.12886          \\
Aesthetics        & 0.12682          \\
Dissimilarity        & 0.12569          \\
Dissimilarity+Arousal      & 0.12497          \\
Aesthetics+Arousal      & 0.13223          \\
Dissimilarity+Aesthetics      & 0.11984          \\
Dissimilarity+Aesthetics+Arousal    & \textbf{0.11657} \\ \bottomrule
\end{tabular}
\caption{The training results of proposed LWC regarding different combination of features. The $Dissimilarity+Aesthetics+Arousal$ gives the best performance.}
\label{tab:training_result_diff_feature}
\end{table}

    The training results of LWC and its baselines are given in Table~\ref{tab:training_result}. Here all the results are given under random seed $s=0$ to ensure the reproducibility. We found that LWC has the best performance on fitting the data with the fewest parameters. This proves LWC is a more cost-effective choice than the general RNN model in the context of evaluating visual-storylines. The RNN and its variants are not better than LWC, which indicates that for small datasets, a carefully designed model may perform better than a general model.
    
    We train LWC using different combinations of features of visual materials as inputs. The results are given in Table~\ref{tab:training_result_diff_feature}, which illustrate that the best results came when all features are used. Additionally, by comparing the results with Table~\ref{tab:results_principle}, we find the training results are aligned with the regression results, where the importance of features are $Informativeness > Aesthetics > Emotion$. Interestingly, we observe that even for LWC using solely $Arousal$ as features, the results are better than RNN and its variants. We tend to agree that the LWC are benefited from the fixed shape of the Wundt Curve, while the training of RNNs is hard to find the best weight because the volume of data is rather small.

    \begin{figure*}[ht]
        \captionsetup[subfigure]{labelformat=empty}
        \begin{subfigure}{0.495\textwidth}
            \includegraphics[scale=0.21,page=10]{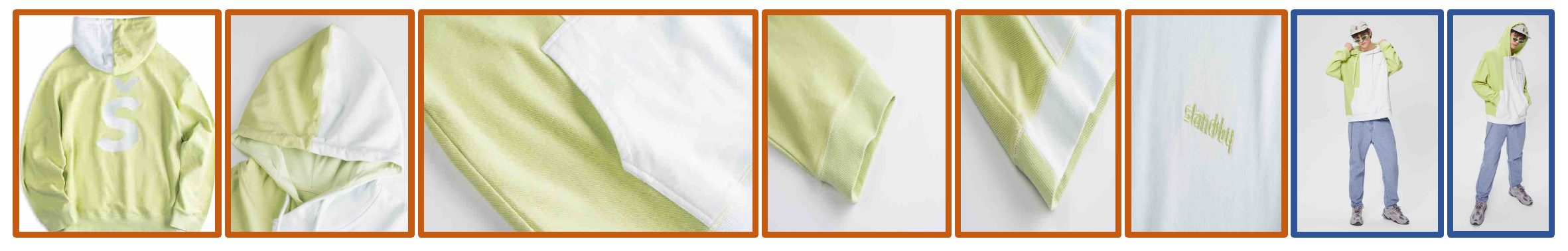}
            \caption{PA \cite{choi2016}}
        \end{subfigure}
        \begin{subfigure}[h]{0.495\textwidth}
            \includegraphics[scale=0.21,page=9]{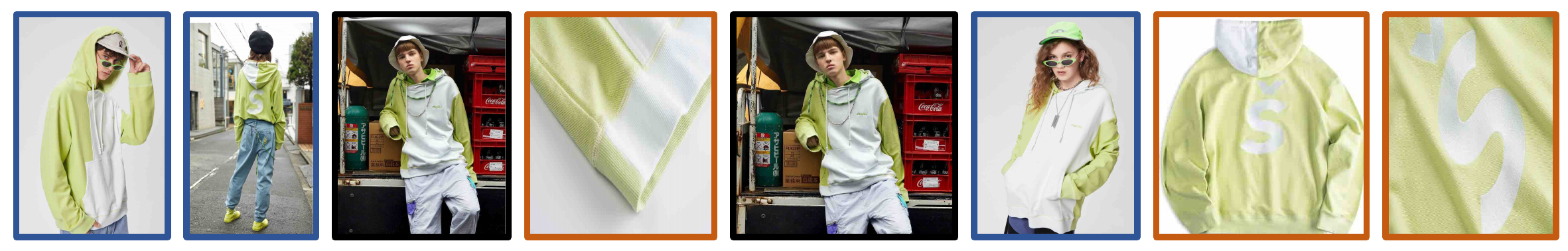}
            \caption{2RNNS \cite{zhong2018}}
        \end{subfigure}
        \begin{subfigure}[h]{0.495\textwidth}
            \includegraphics[scale=0.21,page=11]{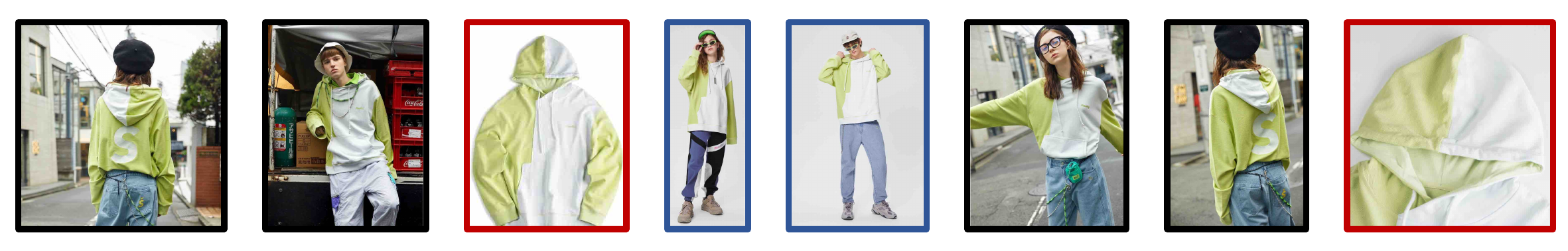}
            \caption{SRNN \cite{sigurdsson2016}}
        \end{subfigure}
        \begin{subfigure}[h]{0.495\textwidth}
            \includegraphics[scale=0.21,page=12]{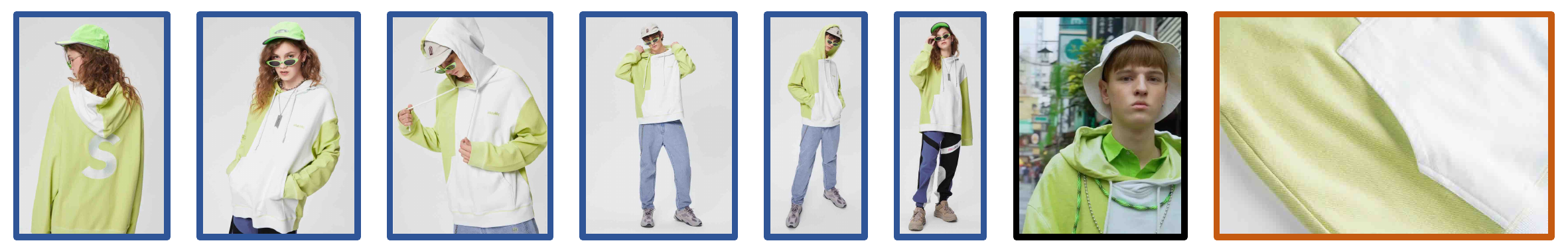}
            \caption{WBP (ours)}
        \end{subfigure}
        \begin{subfigure}{0.495\textwidth}
            \includegraphics[scale=0.21,page=14]{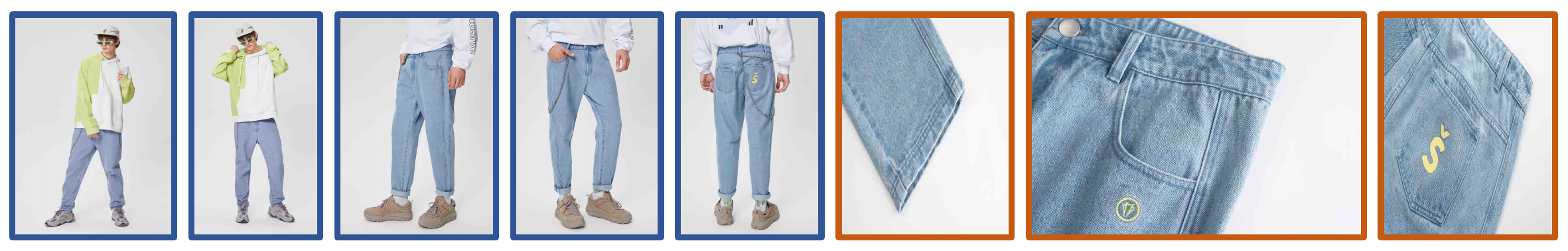}
            \caption{PA \cite{choi2016}}
        \end{subfigure}
        \begin{subfigure}[h]{0.495\textwidth}
            \includegraphics[scale=0.21,page=13]{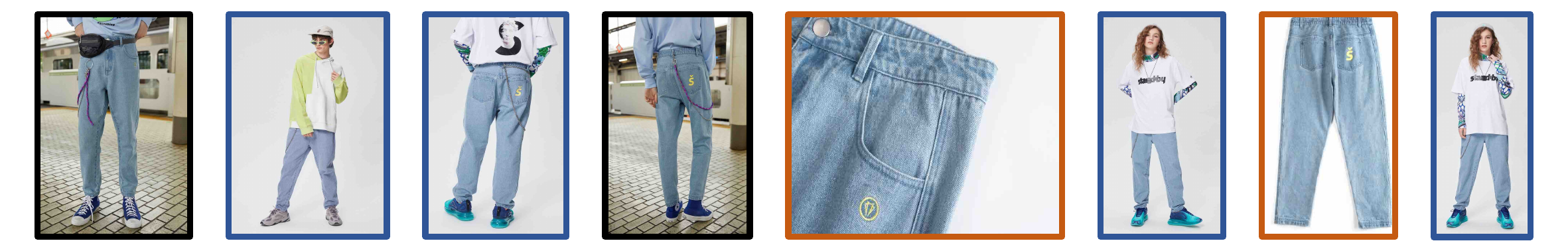}
            \caption{2RNNS \cite{zhong2018}}
        \end{subfigure}
        \begin{subfigure}[h]{0.495\textwidth}
            \includegraphics[scale=0.21,page=15]{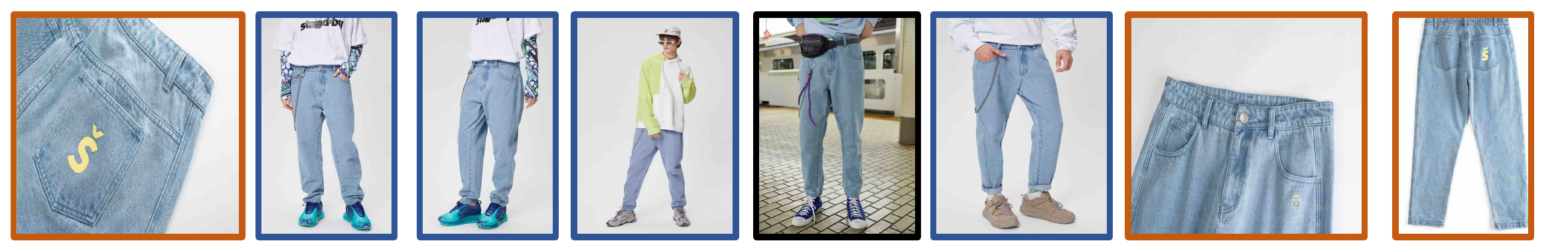}
            \caption{SRNN \cite{sigurdsson2016}}
        \end{subfigure}
        \begin{subfigure}[h]{0.495\textwidth}
            \includegraphics[scale=0.21,page=16]{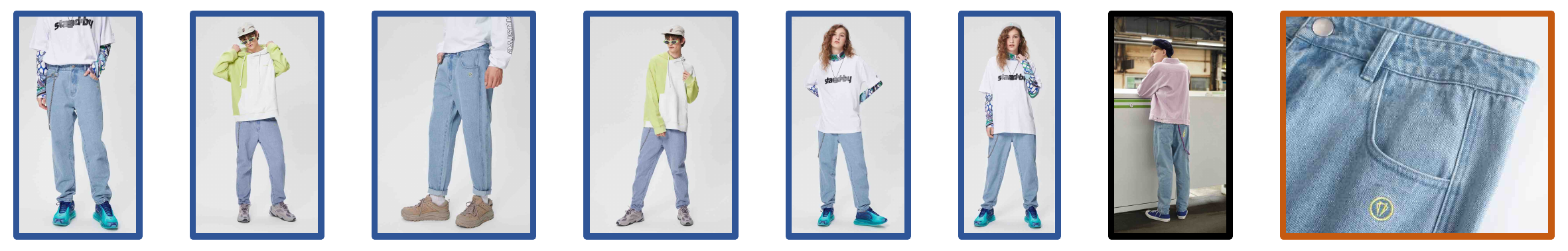}
            \caption{WBP (ours)}
        \end{subfigure}

        \caption{Example visual storylines (best viewed in color), where all visual materials are images. The border color represents the categories of visual materials. Red border: the product detail; blue border: indoor fashion model; black border: outdoor fashion model display. The algorithms used are given below the storyline.}
        \label{img:compari}
    \end{figure*}
    
    The examples of generated visual storyline are shown in Figure~\ref{img:compari}. PA tends to select images similar to the previous ones (especially in the first example), which causes redundancy in terms of information. 2RNNS makes the sequence show dynamism. However, logic is rarely preserved, and for the first case, similar images are selected, which also cause redundancy. SRNN keeps good-looking visual materials. The problem of logic also exists. Our solution is able to put the visual materials with the same topic together, and the materials are placed in a dynamic order. Here in this two examples, three topics are identified, fashion model indoor show, outdoor show and cloth detail show. WBP tends to avoid images like plain detail show, which is identified to have a low attractiveness score and arousal score. What's more, our solution gives the most of the time and space for attractive images like fashion model display, compared with sequences generated by other algorithm. This is because CB tries to reach the maximum perceived persuasiveness (Equation~\ref{equ:dp}) under the constraints of clustering results.
    

    \begin{table}[ht]
        \centering
        \begin{tabular}{ l r r r r }
                \toprule
                Method & PA & 2RNNS & SRNN & WBP \\
                \midrule
                Score & 0.3079 & 0.7161 & 0.3762 & \textbf{0.7869}\\
                \bottomrule
            \end{tabular}
            \caption{The pairwise persuasion scores on all products.}
            \label{tab:pairwise}
    \end{table}
    
    \begin{table}
        \centering
            \begin{tabular}{lrrrrr}
                \toprule
                Shirts & Pants & Skirts & Shoes & Accessories \\
                \midrule
                72.16\% & 14.59\% & 6.49\% & 6.22\% & 0.54\%\\
                \bottomrule
            \end{tabular}
        \caption{Percentage of revenue generated by different categories of products during the November 11th Singles' Day Promotion Taobao clothing sales, 2018 (data source: zhiyitech.cn).}
        \label{img:weight}
    \end{table}
        
    Table~\ref{tab:pairwise} shows the average pairwise persuasion scores on the testing dataset. WBP outperforms the second best algorithm, 2RNNS, by 9.9\%. By jointly considering the sales proportion, shown in Table~\ref{img:weight}, and the scores among different product categories (Figure~\ref{img:cat}), we found that WBP can achieve an expected revenue 12.5\% higher than that achieved by 2RNNS using the following function:
    \begin{equation}
        \bar{R}=\sum_{i \in C}{\frac{P_i(X_i-Y_i)}{Y_i}},
    \end{equation}
    where $C$ is all the categories of products in the test dataset, $X_i$ is the perceived persuasiveness score achieved by WBP for product category $i$, $Y_i$ is the perceived persuasiveness score achieved by 2RNNS for product category $i$, and $P_i$ is the percentage of real-world revenue generated by $i$. 
    
    For PA and SRNN, because video clips are not selected in their generated sequences, the scores are lower. Note that though PA is able to select videos, it prefers images according to the definition of its dynamic function. Figure~\ref{img:wo} shows that when inputs do not contain video clips, WBP achieves 27.9\% higher average perceived persuasiveness score than the the-state-of-art method, SRNN. This indicates that our algorithm can handle the image sequencing well. When inputs contain video clips, WBP outperforms 2RNNS by 2.1\%. From subjects comments, we found that it is because when two storylines have the same long video clips, the previous visual materials in the sequence tend to be overlooked by the test subjects. Figure~\ref{img:cat} and \ref{img:edu_all} illustrate the distributions on perceived persuasiveness for subjects in different categories of products and education level, respectively. It can be observed that for most of the situations, WBP is able to generate visual storylines with a better perceived persuasiveness compared to existing approaches.
    
    \section{Conclusions and Future Work}
    In this paper, we proposed an intelligent solution to generate visual storylines with high perceived persuasiveness for promotional videos. To the best of our knowledge, this is the first work focusing on the persuasiveness of automatically generated videos. The proposed algorithm, WundtBackpack, is composed of Learnable Wundt Curve and Clustering-based Backpacking techniques, which require few training data to achieve high effectiveness. We demonstrate, through real-world experiments, that our algorithm outperforms the best existing approach by achieving 9.9\% higher perceived persuasiveness scores, and 12.5\% higher expected revenue.
    
    Much interesting work is still ahead, including how to decide the optimal display duration of footage, as well as the transition effects between visual materials. These are also important decisions making artificial intelligence (AI) which can make vivid persuasive promotional videos. We will also explore unsupervised learning methods, to make use of the huge amount of promotional videos in E-commerce platforms to improve visual storylines. Since the generated videos are attempting to change people's behaviours or decisions, we will also explore how to incorporate ethical considerations \cite{Yu-et-al:2018IJCAI} into the AI algorithm in order to respect and protect user interest.
    
    In addition, during the experiments, we observed that many people hold different opinions on what constitute good persuasive videos. Some people prefer good-looking models, while others want to see more details of the products. Some people like to see details from the local region of a product first before zooming out to the global view, while others prefer the opposite way. It is difficult to design a fixed storyline to satisfy everyone's preference. Thus, in the future, we will incorporate user prior information into our approach to enable dynamic personalization of generated persuasive videos for different viewers.
\begin{acks}
This research is supported, in part, by the Alibaba-NTU Singapore Joint Research Institute; the Nanyang Assistant Professorship, Nanyang Technological University, Singapore; the National Key R\&D Program No.2017YFB1400100; the Innovation Method Fund of China No.2018IM020200; and the SDNFSC No. ZR2017ZB0420.
\end{acks}
\bibliographystyle{ACM-Reference-Format}
\balance
\bibliography{new_bib}

\end{document}